\documentclass{article}
\usepackage{spconf,amsmath,graphicx,url}
\usepackage{multirow}
\usepackage[
separate-uncertainty = true,
multi-part-units = repeat
]{siunitx}
\usepackage{stfloats}
\usepackage{booktabs}
\usepackage{color,soul}
\usepackage{caption}
\usepackage{graphicx}
\usepackage{hyperref}

\usepackage{fancyhdr}
\title{Investigating Content-Aware Neural Text-To-Speech MOS Prediction \\Using Prosodic and Linguistic Features}
\name{
\begin{tabular}{c}
	Alexandra Vioni$^{\star}$,
	Georgia Maniati$^{\star}$,
	Nikolaos Ellinas$^{\star}$, \\
	June Sig Sung$^{\dagger}$,
	Inchul Hwang$^{\dagger}$,
	Aimilios Chalamandaris$^{\star}$,
	Pirros Tsiakoulis$^{\star}$
\end{tabular}
}

\address{$^{\star}$ Innoetics, Samsung Electronics, Greece \\
	$^{\dagger}$ Mobile eXperience Business, Samsung Electronics, Republic of Korea}

\makeindex

\begin{document}
\ninept
\maketitle
\begin{abstract}
Current state-of-the-art methods for automatic synthetic speech evaluation are based on MOS prediction neural models.
Such MOS prediction models include MOSNet and LDNet that use spectral features as input, and SSL-MOS that relies on a pretrained self-supervised learning model that directly uses the speech signal as input.
In modern high-quality neural TTS systems, prosodic appropriateness with regard to the spoken content is a decisive factor for speech naturalness.
For this reason, we propose to include prosodic and linguistic features as additional inputs in MOS prediction systems, and evaluate their impact on the prediction outcome.
We consider phoneme-level F0 and duration features as prosodic inputs, as well as Tacotron encoder outputs, POS tags and BERT embeddings as higher-level linguistic inputs.
All MOS prediction systems are trained on SOMOS, a neural TTS-only dataset with crowdsourced naturalness MOS evaluations.
Results show that the proposed additional features are beneficial in the MOS prediction task, by improving the predicted MOS scores' correlation with the ground truths, both at utterance-level and system-level predictions.
\end{abstract}
\section{Introduction}
\label{sec:intro}

The emergence of deep learning in text-to-speech (TTS) has greatly broadened its potential, in terms of both speech quality and diversity.
Since neural TTS systems are capable of producing high-quality synthetic speech, focus has shifted to the exploration of style, prosody and expressiveness, in order to build more natural voices. 
The evaluation of synthetic speech heavily relies on subjective mean opinion score (MOS) tests performed by human listeners.

Deep learning-based models for MOS prediction have been recently developed, with the intent to complement or substitute the laborious and expensive listening tests.
Nonetheless, the direction of existing MOS prediction systems is mostly content-agnostic, and only the speech signal or spectrogram is used as input.
But is perceived naturalness not dependent on both what we say and how we say it? Wouldn't an unexpected pause or a prolonged phoneme duration severely affect the perception of naturalness? 
Countless prosodic renditions of a sentence will be natural and acceptable, but certain renditions will always be considered unnatural. 
Is it possible to predict these cases only by considering the spectrogram on the frame-level? 
Such questions have inspired our investigation.

\subsection{Related Work}

Intrusive objective speech quality metrics are not applicable in TTS output, as they require a corresponding original signal to be used as reference for comparison with the generated audio.
Non-intrusive metrics such as P.563 \cite{Malfait2006} are not well-suited for evaluating modern TTS systems either, since they establish speech quality standards aiming to detect compression and transmission artifacts in telecommunications.
Efforts have been made to improve or indicate new metrics that correlate well to the perceived quality of synthetic speech, as in \cite{Falk2008}, using the publicly available data from the Blizzard Challenge (BC) \cite{Black2005}, which include synthetic speech utterances produced by many synthesizers and the corresponding listeners' scores.
Subsequently, researchers have focused on data-driven approaches, training models for utterance-level and system-level evaluation predictions, in which feature extraction is automatically performed by the model itself, e.g., using convolutional neural networks (CNNs), from the spectrograms provided as input to the model \cite{Yoshimura2016, Patton2016, Fu2018}. 

Recent progresses 
have focused on leveraging deep learning architectures and methods, rather than investigating features that can be easily made available to the models.
The release of the large-scale VC Challenge dataset \cite{Lorenzo2018} has set new ground for deep learning-based score prediction.
In MOSNet \cite{Lo2019}, a CNN-BLSTM model predicts the utterance-level MOS for converted speech, by averaging frame-level scores. 
Improvements to this model have been proposed, accounting for listener bias \cite{Leng2021, Huang2021}, and using transfer learning from POLQA prediction models even mixing TTS and VC training data from different listening tests \cite{Mittag2020}.
Recently, self-supervised learning (SSL) networks, pretrained on large quantities of speech data, were introduced to the task by Tseng \textit{et al.} \cite{Tseng21} and have been since shown very effective in generalizing to out-of-domain data \cite{Cooper2021b}.

Their superiority over other models was highlighted in the VoiceMOS Challenge 2022 \cite{Huang2022}, a shared task using common datasets for MOS prediction, where winning teams extended the SSL-MOS baseline to outperform it only by a margin on the third decimal point of the correlation metrics.
Some interesting proposed additions to the baseline include ensembling \cite{Saeki2022, Yang2022}, multi-task learning \cite{Tian2022}, and use of speech recognizers to recreate the phoneme sequence \cite{Saeki2022} or to get ASR evaluations \cite{Yang2022}. 
As the training dataset included VC and TTS systems spanning over a decade \cite{Cooper2021a}, it is unclear if the trained models are able to distinguish between similar systems and utterances, which is a realistic evaluation scenario for TTS researchers.
In prior work \cite{Maniati2022}, we train MOS prediction models with neural TTS-only data on the high-scoring region of the MOS range. 
While SSL-MOS outperforms MOSNet-like models by leveraging its prior knowledge, its utterance-level performance is moderate, indicating room for improvement.


\subsection{Contribution}

In this work, we propose a content-aware evaluation approach for synthetic speech, by modifying several deep learning-based models to include prosodic and linguistic features as inputs for the MOS prediction task.
Relevant works so far have considered only frame-level spectrogram features and have focused on evaluating very diverse systems. 
While the present study is based on these recent approaches, it capitalizes on a new feature space, relevant to the linguistic and prosodic content of the signal, which was not considered in earlier studies. 
We focus on utterance-level evaluation of several similar systems, for reliable MOS prediction on a fine-grained level, which is closer to researchers' needs when evaluating modern TTS, and show that our proposed features are beneficial to the task.


\section{Data}
\label{sec:data}

For our investigation, we utilize the SOMOS dataset \cite{Maniati2022}.
It consists of 20K TTS audio files generated with the LJ Speech voice \cite{LJspeech17} from several Tacotron-like acoustic models and an LPCNet vocoder. 
Each audio was assigned 17-23 naturalness ratings by crowdsourced listeners.

We designed training/validation/test splits\footnote{https://doi.org/10.5281/zenodo.7119399} of approximately 70\%/15\%/15\% of the data for both SOMOS-full and SOMOS-clean sets. 
For each of the validation and test sets, we have selected a small number of unseen categories of listeners, systems and texts based on SOMOS-full, as follows. 
10 listeners were chosen across locales, who met the criteria of having evaluated less than 3,000 samples (size of each of validation/test sets) and submitted less than 50 HITs in total, at least half of which are considered \textit{clean} according to our checks, so that the split is reproducible in SOMOS-clean. 
The resulting unseen listeners for the validation set were 6 US, 2 GB, 2 CA and for the test set 4, 4 and 2 respectively. 
10 systems were selected at random among all models and their samples were included in the respective set. 
50 texts of varying lengths and domains were chosen at random.
While keeping all scores per sample in the same split, we randomly filled the rest of samples and corresponding scores to reach the designated split size. 
To ensure that the score distributions of the splits are similar to the overall dataset score distribution, we have repeated the random sampling process 1000 times with different random seeds and evaluated the candidate splits' distributions using Wasserstein distance as in \cite{Cooper2021b}.
The final split of the dataset was chosen so as to minimize this distance. 

All our experiments are run on the SOMOS-clean subset of the dataset, consisting of 214,745 ratings by 864 listeners, which has been shown as more inherently predictable in \cite{Maniati2022}.

\section{Features}
\label{sec:features}

\subsection{Prosodic}
As \textit{prosodic}, we define the segmental features extracted from the signal that describe the intonation and rhythm of the utterance. 
To extract these, we first input the text sentences of the synthesized corpus to an American English front-end module and obtain a sequence of phones for each utterance.
Then, we derive alignments between each synthesized utterance and its phonetic transcription using the HMM forced-alignment system utilized in \cite{Raptis2016}.
Once the audio is properly aligned, phoneme-level features of duration and pitch (F0) are extracted for the entire dataset. 
F0 is extracted using a standard autocorrelation method \cite{Boersma1993}, followed by interpolation and smoothing of the contour, as in \cite{Vioni2021}.
The phoneme-level F0 feature is calculated as the average of the F0 values for its full duration.

\subsection{Linguistic}

We also employ features derived from text, hence referred as linguistic features.
Syntax plays an important role in the formation of prosodic patterns \cite{Bennett2019, Kohn2018}. Thus we generate part of speech (POS) tags for each token (\textit{pos-tags}), which incorporate syntactic information, output by our front-end module.

BERT \cite{Devlin2019} is a method that learns general-purposed text embeddings based on a bi-directional Transformer encoder, which is trained to predict both randomly masked words and the next sentence in the multitask learning framework.
Its learnt representations contain both semantic and syntactic cues.
To create token embeddings for our corpus (\textit{sem-w-*}), we use the BERT tokenizer and the pretrained base model, and utilize the hidden states of the last layer (\textit{sem-w-*l}), as well as the last 4 layers' hidden states summed (\textit{sem-w-*4s}) and concatenated (\textit{sem-w-*4c}) for each token. 
Each word is mapped to a 768-dimensional vector, except for the latter case where the dimensions are 3,072. 
We have performed the experiments with both cased (\textit{sem-w-c*}) and uncased (\textit{sem-w-u*}) versions of the model.
Additionally, we use the sentence transformers framework \cite{Reimers2019} to extract 768-dimensional sentence embeddings from the pretrained MPNet \cite{Song2020} base model (\textit{sem-utt}). 
MPNet leverages the dependency among predicted tokens through permuted language modeling and takes auxiliary position information as input to make the model see a full sentence and reduce the position discrepancy.

Finally, we create contextual phoneme representations by extracting the encoder outputs of a plain non-attentive Tacotron variant model \cite{Shen2020} trained on US English (\textit{enc-outs}). 
The encoder is identical to our prior work \cite{Ellinas2021}.
Although it is unclear what exactly is learnt in these representations and prosodic cues may be implicitly modeled, we consider these features linguistic, as they are derived by processing the text and its phonetic transcription. 
Since the goal of the encoder during training is to extract robust sequential representations of phonemes, these outputs can be interpreted as containing adequate context of the input phoneme sequence.

\begin{figure*}[t]
	\begin{minipage}[b]{1.0\linewidth}
		\centering
		\centerline{\includegraphics[width=17.5cm]{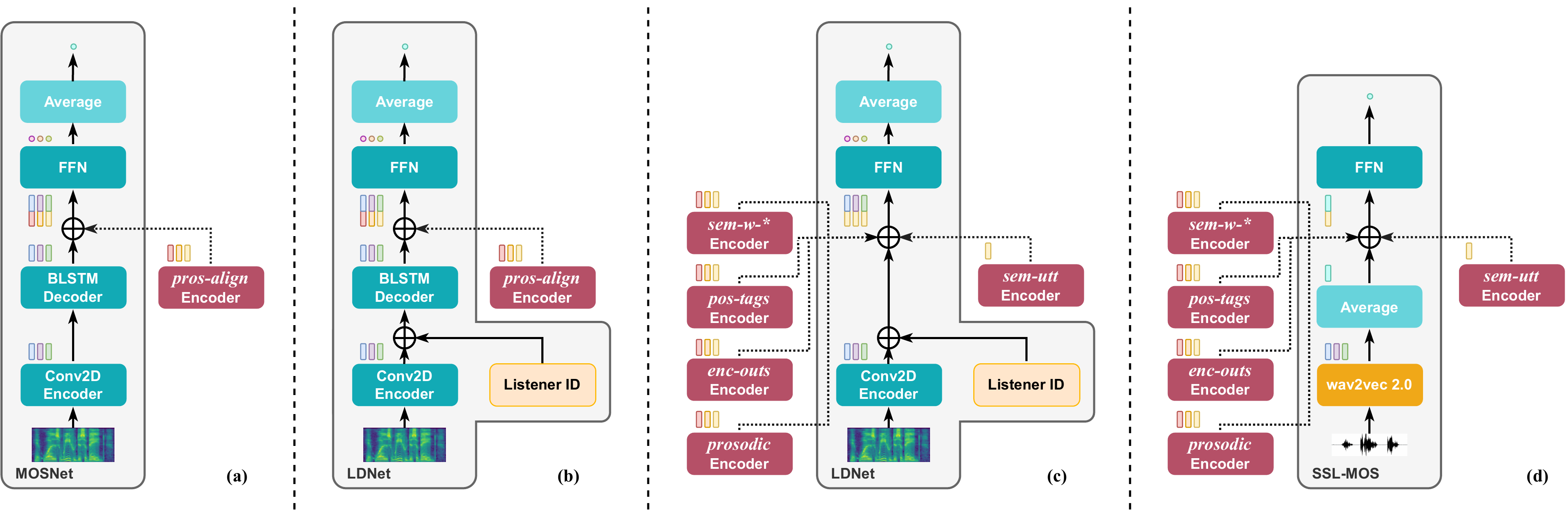}}
	\end{minipage}
	\vspace{-10pt}
	\caption{
		MOS prediction models architecture and proposed additions (in dotted line). 
		Subfigure (a) corresponds to MOSNet, subfigures (b) and (c) to LDNet (modes with and without recurrent decoder) and subfigure (d) to SSL-MOS. Each type of features is tested separately, thus only one feature encoder (dotted line) is used in each experiment.}
	\label{fig:architectures}
	\vspace{-6pt}

\end{figure*}

\section{Experimental Setup}
\label{sec:experiments}

We conduct experiments on three widely used MOS prediction models, namely MOSNet, LDNet and SSL-MOS.
Besides training the models using the SOMOS dataset split, to be used as baselines, we augment each model with each type of additional features separately and re-train them from scratch, to evaluate the features' impact.

The encoder modules which are added to the MOS prediction models according to the features that are being tested, are presented in Table \ref{tab:feature-encoder}.
The phoneme sequence and prosodic features are used in two different ways, the \textit{prosodic} mode, in which we use the raw phoneme and prosodic sequences, and the \textit{pros-align} mode, in which we align these features to the utterance's spectrogram frames, by propagating the phoneme-level values for all the frames of each phoneme.
In both cases, the phoneme labels are mapped into embeddings which are then concatenated to the prosodic features and pass through a single feed-forward layer and a bidirectional GRU (Bi-GRU) layer.
Syntactic \textit{pos-tags} are encoded in a similar architecture.
Tacotron \textit{enc-outs} and word-level BERT embeddings \textit{sem-w-*} are fed to a BLSTM, while utterance-level BERT embeddings \textit{sem-utt} pass through a two-layer feed-forward net.

\vspace{-0.5mm}
\begin{table}[phbt]
	\scriptsize
	\caption{Feature encoder architecture for each type of features}
	\vspace{-2mm}
	\label{tab:feature-encoder}
	\centering
	\begin{tabular}{@{\hspace*{0.5mm}} l | @{\hspace*{4.0mm}} c @{\hspace*{2.0mm}} c @{\hspace*{2.0mm}} c @{\hspace*{2.0mm}} c @{\hspace*{2.0mm}} c @{\hspace*{0.9mm}} }
		\toprule
		\textbf{Features} & \textbf{Embedding} & $\rightarrow$ & \textbf{Feed-forward} & $\rightarrow$ & \textbf{Recurrent} \\
		\midrule
		pros-align & 64-dim & & 128-dim & & 128-dim \\
		prosodic & (phonemes) & & ReLU, dropout & & Bi-GRU \\
		\midrule
		\multirow{2}{*}{enc-outs} & \multirow{2}{*}{-} & & \multirow{2}{*}{-} & & 256-dim \\
		& & & & & BLSTM \\
		\midrule
		\multirow{2}{*}{pos-tags} & \multirow{2}{*}{256-dim} & & 256-dim & & 256-dim \\
		& & & ReLU, dropout & & Bi-GRU \\
		\midrule
		\multirow{2}{*}{sem-utt} & \multirow{2}{*}{-} & & 512-dim & & \multirow{2}{*}{-} \\
		& & & 256-dim & & \\
		\midrule
		\multirow{2}{*}{sem-w-*} & \multirow{2}{*}{-} & & \multirow{2}{*}{-} & & 256-dim \\
		& & & & & BLSTM \\
		\bottomrule
	\end{tabular}
 	\vspace{-5mm}
\end{table}

\subsection{MOSNet}
\label{sec:mosnet}

Lo \textit{et al.} \cite{Lo2019} have proposed MOSNet, a simple architecture that uses speech spectrograms as input and combines convolutional and/or recurrent layers to predict frame-level MOS scores, which are then averaged to extract an utterance-level score.
For our experiments, we implement the combined CNN-BLSTM MOSNet setup in PyTorch, based on the official MOSNet implementation\footnote{https://github.com/lochenchou/MOSNet}.

Aside from training the standard plain MOSNet architecture, we create a model variant which uses the \textit{pros-align} mode of prosodic features as additional inputs.
The outputs of the feature encoder Bi-GRU for each frame are concatenated with the BLSTM frame-level outputs of MOSNet, before passing through the feed-forward MOSNet final layers.

\subsection{LDNet}

LDNet \cite{Huang2021} is based on MOSNet, but its specialized model structure and inference method allow for utilizing individual listener ratings to facilitate listener-dependent modeling.
It comprises an encoder-decoder architecture, where listener information is fed to the decoder, and it includes MeanNet and mean listener additions to acquire the mean score.
For our experiments, we use the official LDNet implementation\footnote{https://github.com/unilight/LDNet} and the mean listener setup.

We train the baseline LDNet model, and variants including the aforementioned prosodic and linguistic features.
Regarding the \textit{pros-align} setup, we use the recurrent decoder option of LDNet, and the frame-level outputs of the feature encoder Bi-GRU are concatenated to the LDNet decoder BLSTM frame-level outputs.

We also test the simple \textit{prosodic} variation, where the prosodic features are not aligned with the spectrogram frames.
For this setup, as well as for all the linguistic features that are tested, we use the feed-forward decoder option of LDNet.
For the \textit{prosodic}, \textit{enc-outs}, \textit{pos-tags} and \textit{sem-w-cl} features, the feature encoder Bi-GRU/BLSTM output corresponding to the last timestep is appended to all LDNet encoder outputs.
Respectively, for the \textit{sem-utt} features, the output of the two-layer feed-forward net is appended to all LDNet encoder outputs.
A single feed-forward layer is added before passing them to the LDNet decoder.

\subsection{SSL-MOS}

The idea of SSL-MOS \cite{Cooper2021b} 
lies on leveraging representations derived from Fairseq\footnote{https://github.com/pytorch/fairseq} SSL speech models, by mean-pooling the models' output embeddings and adding a simple linear fine-tuning output layer.
For our experiments, we use the official implementation\footnote{https://github.com/nii-yamagishilab/mos-finetune-ssl} and employ the wav2vec 2.0 
base model trained on Librispeech.

We train the baseline SSL-MOS model, and variants adding prosodic and linguistic features, again one at a time.
We do not use the \textit {pros-align} setup since wav2vec 2.0 uses the raw audio signal as input and we have not aligned the prosodic features with the wav2vec2.0 outputs' temporal resolution.
For the \textit{prosodic}, \textit{enc-outs}, \textit{pos-tags} and \textit{sem-w-*} features, we define the feature encoder output as the respective Bi-GRU/BLSTM output corresponding to the last timestep, as previously.
For the \textit{sem-utt} features, we define the feature encoder output as the output of the two-layer feed-forward net.
In all cases, we concatenate the feature encoder output with the mean-pooled wav2vec 2.0 output embeddings, before they pass through the feed-forward layer of SSL-MOS.

\subsection{Evaluation Metrics}

Regarding model evaluation, we report the standard MOS prediction metrics, which include the Mean Square Error (MSE) and correlation metrics between true and predicted scores.
Linear Correlation Coefficient (LCC) measures the strength of the linear relationship between the two variables.
Spearman's Rank Correlation Coefficient (SRCC) represents how well the the relationship between the two variables can be described using a monotonic function.
Kendall Tau Rank Correlation (KTAU) is used to measure the ordinal association between the observations of the two variables.
Especially SRCC and KTAU are very useful in practical applications, where monotonicity or high ordinal association between true and predicted scores are requirements for reliable MOS prediction models, in order to accurately compare synthetic utterances or TTS systems, automatically.

\section{Results}
\label{sec:results}

Augmenting the MOSNet model with prosodic features aligned per frame results in a strong improvement on the system-level and a more moderate improvement on the utterance-level (Table~\ref{tab:mosnetldnet}).

\begin{table}[phbt]
	\scriptsize
	\caption{MOSNet \& LDNet utterance-level and system-level results}
	\vspace{-2mm}
	\label{tab:mosnetldnet}
	\centering
	\begin{tabular}{@{\hspace*{0.5mm}} l @{\hspace*{2.0mm}} c @{\hspace*{1.7mm}} c @{\hspace*{1.1mm}} c @{\hspace*{0.9mm}} c @{\hspace*{2mm}} c @{\hspace*{1.7mm}} c @{\hspace*{1.1mm}} c @{\hspace*{0.9mm}} c @{\hspace*{0.4mm}}}
		\toprule
		\multicolumn{1}{c}{} & \multicolumn{4}{c}{\textbf{utterance-level}} & \multicolumn{4}{c}{\textbf{system-level}} \\
		\midrule
		\textbf{Model} (add. features) & \textbf{MSE} & \textbf{LCC} & \textbf{SRCC} & \textbf{KTAU} & \textbf{MSE} & \textbf{LCC} & \textbf{SRCC} & \textbf{KTAU} \\
		\midrule
		MOSNet & 0.271 & 0.486 & 0.470 & 0.325 & 0.081 & 0.667 & 0.679 & 0.479 \\
		MOSNet (pros-align) & \textbf{0.240} & \textbf{0.515} & \textbf{0.498} & \textbf{0.346} & \textbf{0.049} & \textbf{0.781} & \textbf{0.772} & \textbf{0.573} \\
		\midrule
		LDNet & 0.250 & 0.538 & 0.523 & 0.367 & 0.045 & 0.849 & 0.847 & 0.639 \\
		LDNet (pros-align) & 0.230 & 0.542 & 0.521 & 0.367 & 0.037 & 0.841 & 0.837 & 0.630 \\
		LDNet (prosodic) & 0.240 & 0.532 & 0.521 & 0.364 & 0.039 & 0.842 & 0.839 & 0.636 \\
		LDNet (enc-outs) & \textbf{0.223} & \textbf{0.584} & \textbf{0.568} & \textbf{0.401} & \textbf{0.035} & \textbf{0.856} & \textbf{0.856} & \textbf{0.654} \\
		LDNet (pos-tags) & 0.235 & 0.543 & 0.520 & 0.365 & 0.038 & 0.827 & 0.826 & 0.624 \\
		LDNet (sem-utt) & 0.249 & 0.522 & 0.507 & 0.355 & 0.043 & 0.842 & 0.847 & 0.641 \\
		LDNet (sem-w-cl) & 0.241 & 0.554 & 0.529 & 0.370 & 0.036 & 0.838 & 0.834 & 0.627 \\
		\bottomrule
	\end{tabular}
\end{table}

For LDNet, the encoder outputs are the sole features consistently assisting MOS prediction on both utterance and system evaluation (Table~\ref{tab:mosnetldnet}).
Prosodic features were not found to be helpful in this model, regardless of them being aligned with the spectrogram or not.

We also notice that the addition of utterance-level or word-level BERT features does not improve the MOS prediction metrics.
It is not clear whether this is due to a granularity or scope mismatch between the frame-level spectrogram features and the high-level semantic information, due to the absence of ``intermediate" prosodic features, or due to the limited capability of the model and the additional module to capture important semantic cues relevant to the utterances' perceived naturalness.

\begin{table}[phbt]
	\scriptsize
	\caption{SSL-MOS utterance-level and system-level results}
	\vspace{-2mm}
	\label{tab:sslmos}
	\centering
	\begin{tabular}{@{\hspace*{0.5mm}} c @{\hspace*{1.5mm}} l @{\hspace*{2.0mm}} c @{\hspace*{1.7mm}} c @{\hspace*{1.1mm}} c @{\hspace*{0.9mm}} c @{\hspace*{2mm}} c @{\hspace*{1.7mm}} c @{\hspace*{1.1mm}} c @{\hspace*{0.9mm}} c @{\hspace*{0.4mm}}}
		\toprule
		\multicolumn{2}{c}{} & \multicolumn{4}{c}{\textbf{utterance-level}} & \multicolumn{4}{c}{\textbf{system-level}} \\
		\midrule
		& \textbf{Model} (add. features) & \textbf{MSE} & \textbf{LCC} & \textbf{SRCC} & \textbf{KTAU} & \textbf{MSE} & \textbf{LCC} & \textbf{SRCC} & \textbf{KTAU} \\
		\midrule
		\multirow{6}{*}{\rotatebox[origin=c]{90}{batch size 2}}
		& SSL-MOS & 0.194 & 0.648 & 0.633 & 0.452 & 0.027 & 0.892 & 0.897 & 0.713 \\
		& SSL-MOS (prosodic) & \textbf{0.177} & 0.671 & 0.668 & 0.482 & 0.026 & 0.886 & 0.897 & 0.708  \\
		& SSL-MOS (enc-outs) & 0.189 & \textbf{0.679} & 0.670 & 0.485 & 0.032 & 0.906 & \textbf{0.916} & \textbf{0.741} \\
		& SSL-MOS (pos-tags) & \textbf{0.176} & \textbf{0.683} & \textbf{0.675} & \textbf{0.489} & \textbf{0.022} & \textbf{0.911} & \textbf{0.914} & \textbf{0.736}  \\
		& SSL-MOS (sem-utt) & 0.191 & \textbf{0.680} & 0.672 & 0.486 & 0.036 & \textbf{0.910} & 0.911 & 0.731 \\
		& SSL-MOS (sem-w-cl) & 0.182 & \textbf{0.683} & \textbf{0.679} & \textbf{0.492} & 0.033 & 0.896 & 0.910 & 0.731 \\
		\midrule
		\multirow{6}{*}{\rotatebox[origin=c]{90}{batch size 4}}
		& SSL-MOS & 0.209 & 0.671 & 0.663 & 0.477 & 0.050 & 0.907 & 0.909 & 0.729 \\
		& SSL-MOS (prosodic) & 0.189 & 0.669 & 0.656 & 0.472 & \textbf{0.022} & 0.906 & 0.908 & 0.726  \\
		& SSL-MOS (enc-outs) & \textbf{0.178} & \textbf{0.684} & 0.674 & 0.488 & \textbf{0.022} & \textbf{0.911} & \textbf{0.913} & \textbf{0.733}  \\
		& SSL-MOS (pos-tags) & \textbf{0.175} & \textbf{0.680} & 0.672 & 0.486 & \textbf{0.021} & 0.906 & 0.908 & 0.725 \\
		& SSL-MOS (sem-utt) & 0.203 & \textbf{0.687} & \textbf{0.681} & \textbf{0.493} & 0.052 & \textbf{0.911} & \textbf{0.917} & \textbf{0.741} \\
		& SSL-MOS (sem-w-cl) & 0.193 & \textbf{0.684} & \textbf{0.679} & \textbf{0.491} & 0.038 & 0.903 & \textbf{0.913} & \textbf{0.733} \\
		\bottomrule
	\end{tabular}
\end{table}

In SSL-MOS, prosodic features also do not seem to add any valuable information for MOS prediction.
On the contrary, all linguistic features help increase the system's performance, most notably on the utterance-level (Table~\ref{tab:sslmos}).
These changes in the ranking of the best performing features can be attributed to the prior knowledge that is encoded in the SSL-MOS models by wav2vec 2.0. 

We notice that in some cases, the system-level performance is degraded in experiments where utterance-level performance is improved.
This could potentially be attributed to skewed utterance-level predictions that result in slightly lower system-level scores.
Whether prosodic and linguistic features can be leveraged in other ways in SSL-MOS prediction architectures to further improve system-level metrics, remains to be investigated in future research.


It is worth mentioning that the SSL-MOS system is unstable, especially so when it is run with smaller batch size (2 instead of 4).
With the addition of the extra features, we have noticed that training converges earlier, while the system is more consistent in predicted scores, even with a small batch size.
This can be interpreted as efficiency in terms of GPU resources and training time.

\begin{table}[phbt]
	\scriptsize
	\captionsetup{justification=centering}
	\caption{SSL-MOS utterance-level and system-level results regarding variations of word-level BERT embeddings}
	\vspace{-2mm}
	\label{tab:sslmos-semantic}
	\centering
	\begin{tabular}{@{\hspace*{0.5mm}} c @{\hspace*{1.5mm}} l @{\hspace*{1.7mm}} c @{\hspace*{1.7mm}} c @{\hspace*{1.1mm}} c @{\hspace*{0.9mm}} c @{\hspace*{2mm}} c @{\hspace*{1.7mm}} c @{\hspace*{1.1mm}} c @{\hspace*{0.9mm}} c @{\hspace*{0.4mm}}}
		\toprule
		\multicolumn{2}{c}{} & \multicolumn{4}{c}{\textbf{utterance-level}} & \multicolumn{4}{c}{\textbf{system-level}} \\
		\midrule
		& \textbf{Model} (add. features) & \textbf{MSE} & \textbf{LCC} & \textbf{SRCC} & \textbf{KTAU} & \textbf{MSE} & \textbf{LCC} & \textbf{SRCC} & \textbf{KTAU} \\
		\midrule
		\multirow{7}{*}{\rotatebox[origin=c]{90}{batch size 2}}
		& SSL-MOS & 0.194 & 0.648 & 0.633 & 0.452 & 0.027 & 0.892 & 0.897 & 0.713 \\
		\cmidrule(){2-10}
		& SSL-MOS (sem-w-cl) & 0.182 & \textbf{0.683} & \textbf{0.679} & \textbf{0.492} & 0.033 & 0.896 & \textbf{0.910} & \textbf{0.731} \\
		& SSL-MOS (sem-w-c4s) & 0.178 & 0.677 & 0.668 & 0.482 & \textbf{0.020} & \textbf{0.907} & \textbf{0.910} & \textbf{0.728} \\
		& SSL-MOS (sem-w-c4c) & \textbf{0.176} & \textbf{0.680} & 0.668 & 0.482 & \textbf{0.020} & \textbf{0.906} & \textbf{0.910} & \textbf{0.733} \\
		\cmidrule(){2-10}
		& SSL-MOS (sem-w-ul) & 0.179 & 0.673 & 0.668 & 0.483 & 0.025 & 0.884 & 0.898 & 0.711 \\
		& SSL-MOS (sem-w-u4s) & 0.182 & 0.676 & 0.670 & 0.482 & 0.027 & 0.886 & 0.898 & 0.711 \\
		& SSL-MOS (sem-w-u4c) & \textbf{0.177} & 0.677 & 0.669 & 0.482 & \textbf{0.021} & 0.902 & 0.904 & 0.720 \\
		\midrule
		\multirow{7}{*}{\rotatebox[origin=c]{90}{batch size 4}}
		& SSL-MOS & 0.209 & 0.671 & 0.663 & 0.477 & 0.050 & 0.907 & 0.909 & 0.729 \\
		\cmidrule(){2-10}
		& SSL-MOS (sem-w-cl) & 0.193 & \textbf{0.684} & \textbf{0.679} & \textbf{0.491} & 0.038 & 0.903 & \textbf{0.913} & 0.733 \\
		& SSL-MOS (sem-w-c4s) & 0.179 & 0.668 & 0.659 & 0.475 & \textbf{0.021} & \textbf{0.906} & \textbf{0.913} & \textbf{0.736} \\
		& SSL-MOS (sem-w-c4c) & 0.179 & \textbf{0.680} & 0.670 & 0.484 & \textbf{0.021} & \textbf{0.908} & 0.910 & 0.730 \\
		\cmidrule(){2-10}
		& SSL-MOS (sem-w-ul) & \textbf{0.176} & \textbf{0.683} & \textbf{0.677} & \textbf{0.491} & \textbf{0.022} & 0.896 & 0.907 & 0.723 \\
		& SSL-MOS (sem-w-u4s) & 0.185 & 0.663 & 0.653 & 0.469 & 0.024 & 0.900 & 0.907 & 0.725 \\
		& SSL-MOS (sem-w-u4c) & 0.184 & 0.679 & 0.669 & 0.484 & 0.024 & 0.901 & 0.908 & 0.726 \\
		\bottomrule
	\end{tabular}
\end{table}

The cased version of the BERT model performs better across setups, compared to its uncased counterpart.
Cased BERT has been previously shown more effective on named entity recognition and POS tagging tasks, 
which enhances our hypothesis that nuances of syntax are of importance for perceived naturalness and thus for predicting MOS.
In terms of the different approaches for creating token embeddings from BERT representations, the states of the last hidden layer bear consistently the best scores for this task.


\section{Conclusions and Future Work}
\label{sec:conclusions}

We have investigated content-aware MOS prediction by augmenting existing models with several prosodic and linguistic features, including phoneme-level F0 and duration, Tacotron encoder outputs, POS syntactic tags and BERT embeddings.
Our experimental results show that the proposed additional features can be leveraged in the MOS prediction task to improve the systems' performance.
Aligned prosodic features are valuable in MOSNet, Tacotron encoder outputs were found to be beneficial in LDNet, and the higher-level linguistic features appear to slightly improve SSL-MOS predictions.

In future work, we plan to experiment with diverse model architectures as well as alternative ways of encoding the content information to the models.
In addition, we aim to examine the impact of the additional features in the generalization capabilities of the models based on self-supervised speech representations.
At the same time, we aspire to investigate practical uses of MOS prediction models as complementary tools to comparatively evaluate neural TTS systems.


\bibliographystyle{IEEEbib}
\bibliography{refs}

%


\end{document}